\documentclass{osa-article}

\journal{osajournal}
\articletype{Research Article}
\begin{document}

\title{Optimal Laguerre-Gaussian Modes for High-Intensity Optical Vortices}

\author{Andrew Longman,\authormark{1,*} and Robert Fedosejevs\authormark{1}}

\address{\authormark{1}Department of Electrical and Computer Engineering, University of Alberta, Edmonton, Canada, T6G1R1}

\email{\authormark{*}longman@ualberta.ca} %% email address is required

%%%%%%%%%%%%%%%%%%% abstract %%%%%%%%%%%%%%%%

\begin{abstract}
With increasing interest in using orbital angular momentum (OAM) modes in high-power laser systems, accurate mathematical descriptions of the high-intensity modes at focus are required for realistic modelling. In this work, we derive various high intensity orbital angular momentum focal spot intensity distributions generated by Gaussian, super-Gaussian, and ideal flat-top beams common to high-power laser systems. These intensity distributions are then approximated using fitted Laguerre-Gaussian basis functions as a practical way for describing high-power OAM beams in theoretical and numerical models.
\end{abstract}

%%%%%%%%%%%%%%%%%%%%%%%%%%  body  %%%%%%%%%%%%%%%%%%%%%%%%%%
\section{Introduction}
It is well understood that light carries angular momentum. Pedagogically, this is introduced using the concept of circular polarization described classically by the rotation of the electric field vector about a point, or quantum mechanically as photon spin \cite{griffiths_schroeter_2018}. A single photon has a spin angular momentum of $S=\sigma\hbar$ where $\sigma=\pm 1$ depending on its polarization. The total spin angular momentum carried by a circularly polarized high-power laser (HPL) beam can be substantial and its interaction with plasma can lead to interesting effects such as the inverse Faraday effect \cite{Najmudin:01,Naseri:10}, and enhanced particle acceleration and guiding methods \cite{seryi:16}. 

It is possible for light to carry an orbital angular momentum (OAM) in addition to its spin angular momentum. Electromagnetic OAM can be described classically as a beam propagating with a helical wavefront, or quantum mechanically as a single photon with a helical phase in its wavefunction \cite{Yao:11,Padgett:17}. The OAM carried by each photon is directly proportional to $\hbar$ by an azimuthal quantum number $\ell$. Summing over the photon OAM and spin states in a beam of $N$ photons gives the total angular momentum \cite{Allen:92}:
\begin{equation} \label{eq0}
    J=(\sigma+\ell)N\hbar
\end{equation}
The contribution of both OAM and spin may be constructive or destructive such that a circularly polarized beam can have a net zero total angular momentum, or a linearly polarized beam can contain angular momentum via OAM alone. As the OAM of a photon is theoretically unbounded, the ability for a single photon to carry multiple units of angular momentum has drawn large interest, particularly in the fields of communication multiplexing \cite{Gibson:04}, particle trapping \cite{opticaltrap:97}, and laser-plasma interactions \cite{VieiraNC:16,Zhang:14,MendoncaPOP:14,VieiraPRL:18,Pukhov:18}.

Interest in high-power and high-intensity OAM beams has been largely explored theoretically and in particular, its interaction with plasma for enhancement of particle guiding \cite{Zhang:16,VieiraPRL:14}, high-harmonic generation \cite{Rego:19,MendoncaPOP:15}, and magnetic field generation \cite{Shi:18,Ali:10}. Experimental demonstration of high-intensity OAM beams has been hindered due to difficulties associated with generating OAM beams in HPL systems. Typical generation methods used in low-power applications such as cylindrical mode converters \cite{Allen:92,Beijersbergen93}, spatial light modulators \cite{Forbes:16}, and spiral phase plates \cite{Sueda:04,Beijersbergen94} cannot be used in HPL's due to their low damage thresholds and non-linearites introduced by transmitting ultra-short laser pulses through optical materials. 

Recent advances have enabled the generation of high-power OAM beams for exploring these new and interesting phenomena \cite{Denoeud:17,Brabetz:15,Leblanc:17}. A robust and pulse-shape preserving method has been more recently demonstrated using an off-axis spiral phase mirror (OASPM) which could practically introduce OAM to many HPL facilities \cite{Longman:20}. The OASPM imprints a helical wavefront onto the near-field of the laser after pulse amplification and compression via a nano-structured staircase on its surface. The helical beam is then focused to a high-intensity donut mode in the far-field as a result of the phase singularity on the beam axis. The intensity profile of the donut is highly dependant on the symmetry, and the pitch of the helical wavefront of the near-field profile. Typically, donut modes are described in the paraxial regime using a Laguerre-Gaussian (LG) basis set \cite{Allen:92}, of which many modes are often required to accurately represent the diffraction theory of the focal spot. Optimal selection of the LG basis set parameters allows for a single LG mode to contain the majority of the beam energy at focus \cite{Longman:17}. 

With the introduction of experimental high-intensity OAM beams, it has become necessary to understand and model the focal spots further such that an accurate comparison can be made when studying the enhancements potentially introduced by OAM. In this work, we find the intensity distribution of OAM modes at focus generated by Gaussian, super-Gaussian, and ideal flat-top beams via Fraunhofer diffraction. Optimal LG basis sets are then found to model each of the far-field (or focal spot) intensity profiles. From these optimized LG modes, it is possible to construct Maxwell consistent electromagnetic field distrutions of the OAM beam at focus using an appropriate method \cite{Peatross:17}.  

\section{Far-field diffraction of OAM beams}
In this paper, we are concerned with the problem of the far-field intensity profile generated by focusing a beam of a given near-field intensity profile imprinted with a smooth helical wavefront. We thus assume the OAM generating optic imprints a perfectly helical wavefront into the beam of topological charge $L$. We will only consider beams of integer topological charge, that is, the helical pitch of the wavefront $Q$ will be an integer multiple of the wavelength of the laser $Q = L\lambda$. Using this constraint, we are able to restrict our LG basis sets required to describe the OAM focal spot to having only one azimuthal mode number $\ell=L$ \cite{Longman:17}. In principal it is possible to generate non-integer topological charge OAM beams \cite{Gotte:08}, in which the generated OAM focal spot would require multiple LG azimuthal modes to accurately describe the focus. Typically, having non-integer topological charges is not desirable for high intensity studies as it can lead to focal spot asymmetries and an overall reduction in peak intensity. 

An assumption of this analysis will be that of a monochromatic wave. Modern ultra-fast lasers capable of generating ultra-high intensities contain a considerable bandwidth, often surpassing 50nm for titanium-sapphire based systems \cite{backus:98}. The effect of such a bandwidth on the conversion efficiency of spiral phase optics to an LG mode has however been shown to be minimal \cite{Longman:17}, and hence we do not consider its contribution in this work.

We consider a monochromatic, azimuthally symmetric near-field laser amplitude profile with an arbitrary radial function $U(r)$. After imprinting it with a helical phase of integer topological charge $\ell$, we can write the near-field amplitude as,

\begin{equation} \label{eq1}
    U(r,\theta,\ell)=U(r)e^{i\ell\theta}
\end{equation}
Here, $r$ and $\theta$ represent the radius and azimuthal angle in the near-field plane. If the beam has a central wavelength $\lambda$ and is focused by an achromatic optic of focal length $f$, the intensity profile at focus can be found via the Fraunhofer approximation utilizing the 2D Fourier transform in polar coordinates \cite{goodman2005introduction}:

\begin{equation} \label{eq2}
    I(\rho,\phi,\ell) \propto \left|\iint \frac{U(r)}{\lambda f}exp\left(\frac{-i2\pi}{\lambda f}r\rho cos(\theta-\phi) + i\ell\theta \right)rdrd\theta\right|^2
\end{equation}
We use the variables $\rho$ and $\phi$ as the polar coordinates in the far-field plane. Following the standard procedure to integrate this function, we integrate the azimuthal function first by making a change of variables such that $\psi=\theta-\phi$ giving the azimuthal integral:
\begin{equation} \label{eq3}
    e^{i\ell\phi} \int_{0}^{2\pi}exp\left(\frac{-i2\pi}{\lambda f}r\rho cos(\psi) + i\ell\psi\right)d\psi
\end{equation}
Using the identity for the $\ell^{th}$ order Bessel function of the first kind \cite{BornWolf:1999:Book},
\begin{equation} \label{eq4}
    J_{\ell}(a)=\frac{i^{-\ell}}{2\pi}\int_{0}^{2\pi}e^{-iacos(\psi)}e^{i\ell\psi}d\psi,
\end{equation}
we can solve the azimuthal integral in a closed form.

Combining Eq. \ref{eq2} with this identity, we can find the far-field intensity profile at focus for an arbitrary near-field spatial profile carrying OAM:
\begin{equation} \label{eq5}
    I(\rho,\ell)\propto\left|\frac{2\pi}{\lambda f} \int_{0}^{\infty}U(r)J_{|\ell|}\left(\frac{2\pi r\rho}{\lambda f}\right)rdr\right|^2
\end{equation}
We note that this is simply the Hankel transform of order $\ell$ and scaling factor $2\pi\rho/\lambda f$ of the initial near-field profile. In the following sections we explore the solutions of this equation for common HPL near-fields and find the corresponding optimal LG basis set to describe each solution. 

\section{ Gaussian OAM beams}
We first consider a Gaussian near-field amplitude profile. The near-field of which is given in the following form,
\begin{equation} \label{eq6}
    U(r) = U_0exp\left(-\frac{r^2}{R_0^2}\right)
\end{equation}
Here, $R_0$ is the characteristic near-field beam radius and $U_0$ is the peak field amplitude. The power contained in the near-field profile can be found by, 
\begin{equation}
      P \propto\iint U^*(r)U(r)rdrd\theta = \frac{|U_0|^2 \pi R_0^2}{2}
\end{equation}
Throughout this paper we assume the same drive laser power, regardless of its radial shape. As a result the peak field amplitude will change profile to profile, but the total power remains constant. 
Considering a beam with no OAM ($\ell=0$) and assuming no power losses in the focussing of the beam, the far-field intensity profile then takes the familiar form,
\begin{equation} \label{eq7}
    I(\rho,\ell=0) = I_0e^{-2\xi^2}
\end{equation}
Here, $I_0$ is the peak intensity which is maximal on axis, and we introduce a dimensionless transverse spatial variable $\xi$ given by,
\begin{equation}
    \xi=\frac{\rho}{w_0}
\end{equation}
where the Gaussian beam waist is defined as:
\begin{equation} \label{eq8}
    w_0=\frac{\lambda f}{\pi R_0}
\end{equation}
For the $\ell\neq 0$ solution, we are able to solve the Hankel transform analytically using the identity \cite{gradshteyn2007},
\begin{equation} \label{eq9}
    \int_{0}^{\infty}e^{-\alpha r^2}J_{n}(\beta r)rdr=\frac{\sqrt{\pi}\beta}{8\alpha^{\frac{3}{2}}}exp\left(\frac{-\beta^2}{8\alpha}\right)\left[I_{\frac{n-1}{2}}\left(\frac{\beta^2}{8\alpha}\right)-I_{\frac{n+1}{2}}\left(\frac{\beta^2}{8\alpha}\right)\right]
\end{equation}
After some algebra we arrive at the generalization of Eq.\ref{eq7} to include OAM \cite{Kotlyar:05,Kotlyar:06}, 
\begin{equation} \label{eq10}
    I(\rho,\ell)=I_0\frac{\pi}{4}\xi^2e^{-\xi^2}\left[I_{\frac{\ell-1}{2}}\left(\frac{\xi^2}{2}\right)-I_{\frac{\ell+1}{2}}\left(\frac{\xi^2}{2}\right)\right]^2
\end{equation}
Here, $I_{\nu}(x)$ is the modified Bessel function of the first kind. Eq.\ref{eq10} has been normalized such that it contains the same total power as the results in Eq's.\ref{eq6} and \ref{eq7}, and written such that the value of $I_0$ corresponds to the same value as in Eq.\ref{eq7}. The first four solutions of Eq.\ref{eq10} are plotted in Fig.1. As the peak intensity for an OAM beam is off-axis, its peak intensity $I_p$ is a fraction of the value of $I_0$.  The ratio of the first four OAM peak intensity values to the $\ell=0$ peak intensity $I_0$ are given in column 2 of Table 1.

\begin{figure}[ht!] \label{fig1}
\centering\includegraphics{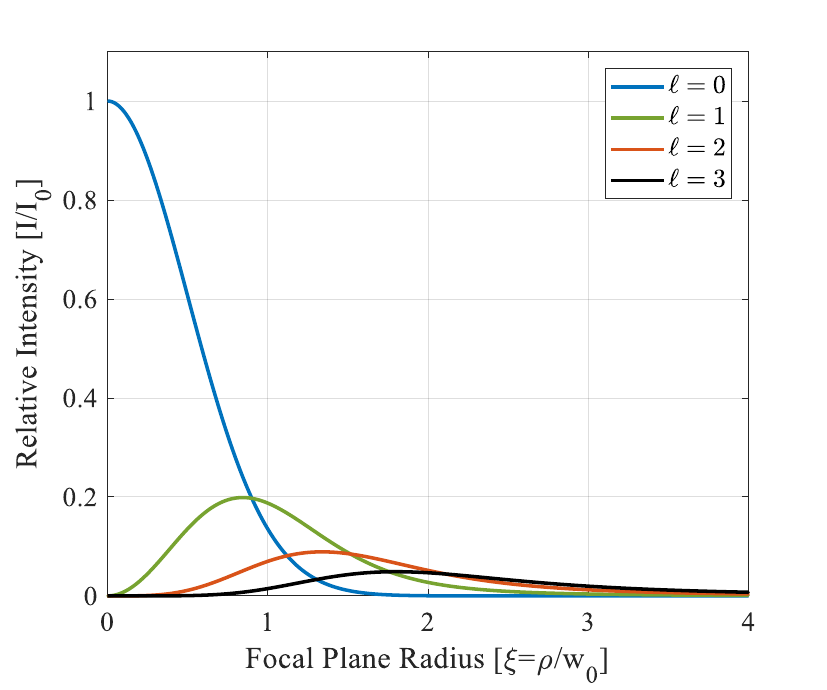}
\caption{Diffraction theory far-field line-outs for the first four topological charge OAM beams driven by a Gaussian near-field beam.}
\end{figure}

\begin{table}\label{table1}
\centering 
\caption{
		Properties of Gaussian and flat-top driven OAM beams with corresponding Laguerre-Gaussian function fitting parameters. Columns 2-4, Peak intensity values and LG fitting parameters ($\gamma_G, \eta_G$) for OAM beams driven by Gaussian near-field beams. Columns 5-7, Peak intensity values and LG fitting parameters ($\gamma_F, \eta_F$) for OAM beams driven by flat-top near-field beams. See text for details of fitting parameters. 
	}
	
		\begin{tabular}{c c c c c c c}
		\hline
			\textrm{$\ell$}&
			\textrm{$I_{Gp}/I_0$}&
			\textrm{$\gamma_G $}&
            \textrm{$\eta_G $}&
            \textrm{$I_{Fp}/I_{0} $}&
            \textrm{$\gamma_F $}&
			\textrm{$\eta_F$} \\
			\hline
			0 & 1.000 & 1.000 & 1.000 & 0.500 & 0.771 & 0.841\\
			1 & 0.199 & 0.837 & 0.771 & 0.129 & 0.576 & 1.061 \\
			2 & 0.089 & 0.748 & 0.589 & 0.073 & 0.508 & 1.047 \\
			3 & 0.049 & 0.681 & 0.470 & 0.049 & 0.465 & 1.010\\
			4 & 0.030 & 0.629 & 0.391 & 0.036 & 0.432 & 0.980\\
			\hline
		\end{tabular}
\end{table}

While Eq.\ref{eq10} is a closed form of the exact intensity distribution at focus, it is useful to describe additionally the distribution in terms of simpler LG modes. The LG basis set can be substituted as an approximation to this equation by suitably choosing the LG mode parameters. We write the intensity of a single LG mode in the far-field in the following form,

\begin{equation} \label{eq11}
    I_{LG}(\rho,\ell,p)=I_0\frac{\eta p!}{(p+|\ell|)!}\left(\frac{w_{0}}{w_{LG}}\right)^2\left[\frac{\rho\sqrt{2}}{w_{LG}}\right]^{2|\ell|}exp\left(\frac{-2\rho^2}{w_{LG}^2}\right)\left[L_p^{|\ell|}\left(\frac{2\rho^2}{w_{LG}^2} \right) \right]^2
\end{equation}
again maintaining the total power in each mode and the value of $I_0$ to be consistent with that in Eq's.\ref{eq7} and \ref{eq10}. The azimuthal mode number $\ell$ is directly proportional to the OAM per photon in the beam while the radial mode number $p$ indicates the number of concentric donut rings ($p+1$) introduced by the generalized Laguerre polynomial $L_p^{|\ell|}$. For the purpose of this study, we will only focus on fitting LG modes with $p=0$ as we are only interested in an approximate first order fit. We introduce the mode scaling factor $\eta$, and define a new beam waist parameter $w_{LG}$ which is proportional to the beam waist of the Gaussian beam through the relationship,
\begin{equation} \label{eq12}
    w_0=\gamma w_{LG}
\end{equation}
The beam waist ratio $\gamma$ has been introduced previously \cite{Longman:17}, and is used as an optimization parameter for the beam waist and to optimize the mode scaling factor $\eta$. The mode scaling factor is directly proportional to the total power in each LG mode given by,
\begin{equation} \label{eq12a}
    P \propto \int_0^{2\pi}\int_0^{\infty}I_{LG}\rho d\rho d\phi=\frac{I_0\eta \pi w_0^2}{2}
\end{equation}In Table 1, we denote the scaling factor and beam waist ratio of a Gaussian driven OAM beam as $\eta_G$ and $\gamma_G$ respectively. This notation is later removed in Tables 2 and 3 as distinction between the drive beams is given by the super-Gaussian parameter $n$.  

We can approach the LG beam waist and scaling factor optimization problem from two different viewpoints. The first is to find a LG basis set that maximizes the total power in the focal spot into a single LG mode. This approach has been previously used \cite{Longman:17} and works well for $l$ numbers less than approximately 5, however for beams carrying more OAM many LG $p$ modes are required to describe the beam accurately. Using this method to describe non-Gaussian near fields becomes yet more complicated requiring further $p$ modes making the method unsuitable for this problem.

We therefore opt for a simple fitting approach where we find a single LG mode that is fitted to match the far-field peak intensity (both radius and magnitude) for each OAM mode given the input near-field profile, and topological charge $\ell$. The normalized radial position of the peak intensity of a $p=0$ LG mode given by Eq.\ref{eq11} is found to be,
\begin{equation} \label{eq13}
    \xi_{max}=\sqrt{\frac{\ell}{2\gamma^2}}
\end{equation}
This relation is then matched to the normalized radial position of the peak intensity of Eq.\ref{eq10}. The peak positions of the diffraction result must be found by a numerical method and once found, can be used to solve for $\gamma$ in Eq.\ref{eq13}. To match the peak intensity of our fitted LG beam to the diffraction far-field result, we use the relation of the peak intensity of the $\ell^{th}$ order, $p=0$ LG mode,
\begin{equation} \label{eq14}
\frac{I_{LG}(\ell)}{I_0} = \frac{|\ell|^{|\ell|}e^{-|\ell|}\gamma^2\eta}{|\ell|!}
\end{equation}
The optimal scaling factor $\eta$ can be found by dividing the peak intensity of Eq.\ref{eq10} by the result of Eq.\ref{eq14} to match the peak intensity value,
\begin{equation} \label{eq15}
\eta = \frac{I(\xi_{max},\ell)|\ell|!}{|\ell|^{|\ell|}e^{-|\ell|}\gamma^2}
\end{equation}
The results of the first four $\ell$ values of $\gamma_G$ and $\eta_G$ are given in  columns 3 and 4 in Table 1. The results of the optimal LG modes both for the $\ell = 1$ and $\ell=2$ modes are plotted and compared to the diffraction theory result in Fig.2 and also a non-optimized ($\eta=1, \gamma=1$) $\ell=1$ LG mode. 
\begin{figure}[ht!] \label{fig2}
\centering\includegraphics{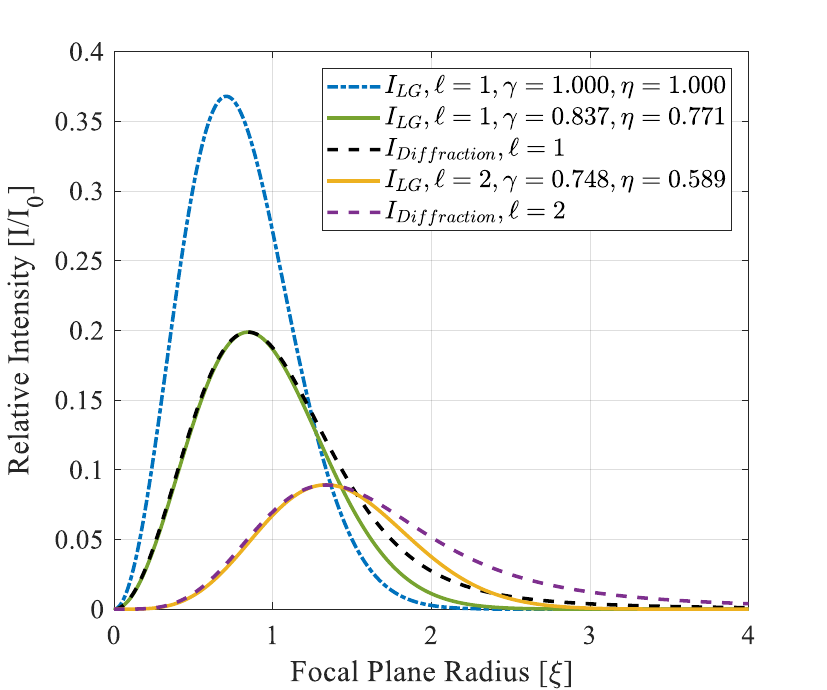}
\caption{Laguerre-Gaussian beams fitted to the exact diffraction result (dashed) for OAM beams driven by Gaussian near-field beams for both $\ell=1$ and $\ell=2$ topological charges. The dash-dot blue line represents a non-optimized LG mode.}
\end{figure}

We note that by modifying $\eta$ to have a non-unitary value, we are under/over-estimating the power in the focal spot when using a single LG mode. For most numerical applications of high-intensity OAM beams, the peak intensity of the focal spot is more critical for modelling than the total focal spot power. From Fig.2, it is clear to see that the LG mode intensity diverges from the diffraction theory for radii larger than the peak intensity as a result of the reduced power ($\eta<1$) in each mode. For radii inside the peak intensity, the LG mode approximates the intensity very well. We believe that this model is suitable for simulating and modelling high-intensity OAM focal spots. In previous publications, this fitting of the LG mode to the diffraction result has not been implemented and as such, the beam waist ratio, and the scaling factor have both been assumed to be unity in the models $\eta=1,\gamma=1$. As such, these lead to a very poor representation of the real OAM focal spot as can be seen in Fig.2.

\section{Flat-top OAM Beams}
While the Gaussian near-field is often assumed in modelling and OAM conversion calculations, HPL's are not typically described by a Gaussian in the near-field as a result of the large transverse extent of the laser amplifiers. Often they can be better described in the near-field to be an ideal flat-top beam where they maintain a constant intensity across a disc of a given radius $R_F$. We can model this near-field with the following formula, 
\begin{equation} \label{eq16}
U(r)=\frac{U_0}{\sqrt{2}} circ\left(\frac{r}{R_F}\right)
\end{equation}where, 
\begin{equation} \label{eq17}
circ\left(\frac{r}{R_F}\right)= \begin{pmatrix} 1, & r<R_F \\ \frac{1}{2}, & r=R_F \\ 0, & otherwise \end{pmatrix} 
\end{equation}
The value of the peak field amplitude $U_0$ is the same as in the Gaussian case but is normalized by a factor of $1/\sqrt{2}$ to ensure the total power in the near-field beam remains the same as in the Gaussian case. This also requires that the near-field beam radius be the same in the Gaussian case and in the flat top case: $R_0 = R_F$.

As before, we first consider a beam without OAM $(\ell=0)$, and after taking the Hankel transform of Eq.\ref{eq16} we yield the well known Airy spot at focus,
\begin{equation} \label{eq18}
I_F(\rho,\ell=0)=\frac{I_{0}}{2}\left[\frac{J_1\left(2\xi\right)}{\xi}\right]^2
\end{equation}Maintaining the same value of $I_0$ as in the Gaussian case, the peak intensity on axis is now found to be $I_{0}/2$ as a result of the power spread into the Airy rings. The beam waist is given by the same definition as in the Gaussian case as the near-field beam radius is the same, as is the wavelength and the focal length of the optic. 

A solution to the $\ell^{th}$ order Hankel transform of Eq. \ref{eq16} becomes more challenging, for which analytical solutions only exist when $\ell$ is even. In general, there seems to be no closed form to the integral, but it is possible to represent the solution as a generalized hypergeometric function ${}_{1}F_2(a;b;z)$ \cite{gradshteyn2007,Kotlyar:06b}:
\begin{equation} \label{eq21}
I_F(\rho,\ell)=\frac{I_{0}}{2}\left|\frac{\xi^\ell}{\ell!\left(\frac{\ell}{2}+1\right)}{}_{1}F_2\left(\frac{\ell}{2}+1;\ell+1,\frac{\ell}{2}+2;-\xi^2\right)\right|^2
\end{equation}Again, the result is normalized such that it contains the same power as previous solutions and as such, the value of $I_0$ is the same as that in the Gaussian focus. The intensity line-outs of Eq.\ref{eq21} for the first four topological charges are given in Fig. 3. The ratios of the peak intensity of each OAM mode compared to $I_0$ is given in column 4 of Table 1. Comparing the flat-top focal spot intensities to those generated by a Gaussian near field in column 2 in Table 1 indicates that the relative peak intensity between the OAM focal spot and the $\ell=0$ focal spot is improved, particularly with beams carrying more OAM. For instance, the relative peak intensity ratio of an $\ell=3$ focal spot generated by a flat-top beam to the $l=0$ peak intensity ($0.049/0.5$), is double to that of the $\ell=3$ peak intensity ratio ($0.049/1$) generated by a Gaussian beam. This increase in relative intensity is likely a result of diffractive effects from the circular aperture function, where the intensity peaks of the Airy rings are overlapped with the intensity peaks of the OAM donut modes. We also find that the peak intensity radius of a flat-top driven OAM focal spot is larger than that driven by a Gaussian near-field.  

\begin{figure}[ht!] \label{fig3}
\centering\includegraphics{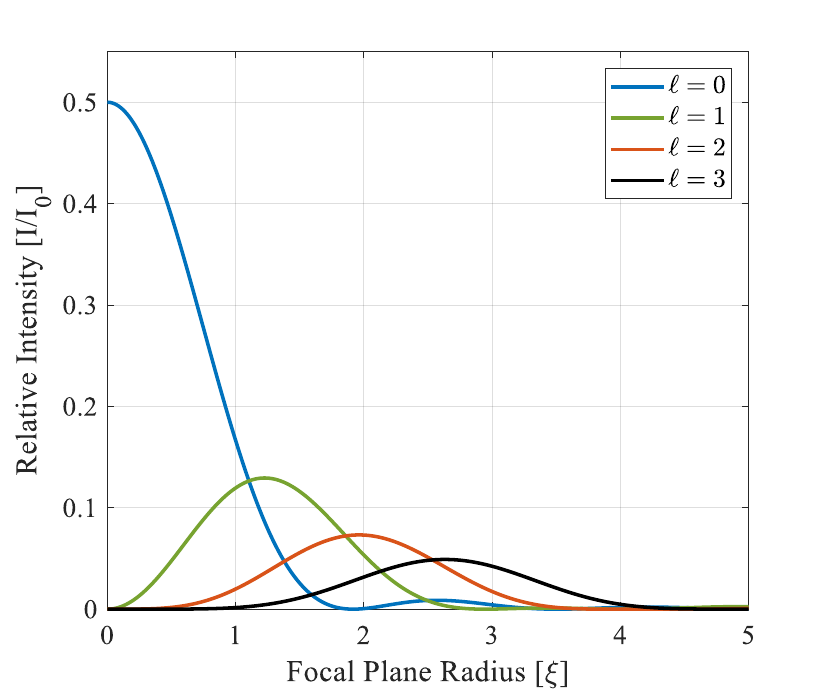}
\caption{Diffraction theory far-field line-outs for the first four topological charge beams driven by a flat-top near-field beam.}
\end{figure}

The form of Eq.\ref{eq21} is not easy to manipulate in beam calculations, so we once again seek an optimal LG mode that can describe each these focal spots. Utilizing the same formulas given in Eq's.\ref{eq13} and \ref{eq15}, we can find the optimal beam waist ratio for the flat-top driven OAM beam $\gamma_F$, and the corresponding mode scaling factor $\eta_F$ where the $F$ subscript indicates a flat-top driver. These results for the first five topological charges are given in columns 6 and 7 in Table 1. The first three fitted LG modes are plotted in Fig.4 with the corresponding fit parameters and show good representation of the diffraction theory far-fields.  

A surprising result of the flat-top driven OAM beams is that a single LG mode represents well the far-field for a broad range of topological charges. The mode scaling factor $\eta_F$ can be seen to be close to unity for the first 5 topological charges given the optimal beam waist ratio $\gamma_F$, unlike the results of the fitted LG modes for Gaussian driven OAM beams. 

\begin{figure}[ht!] \label{fig4}
\centering\includegraphics{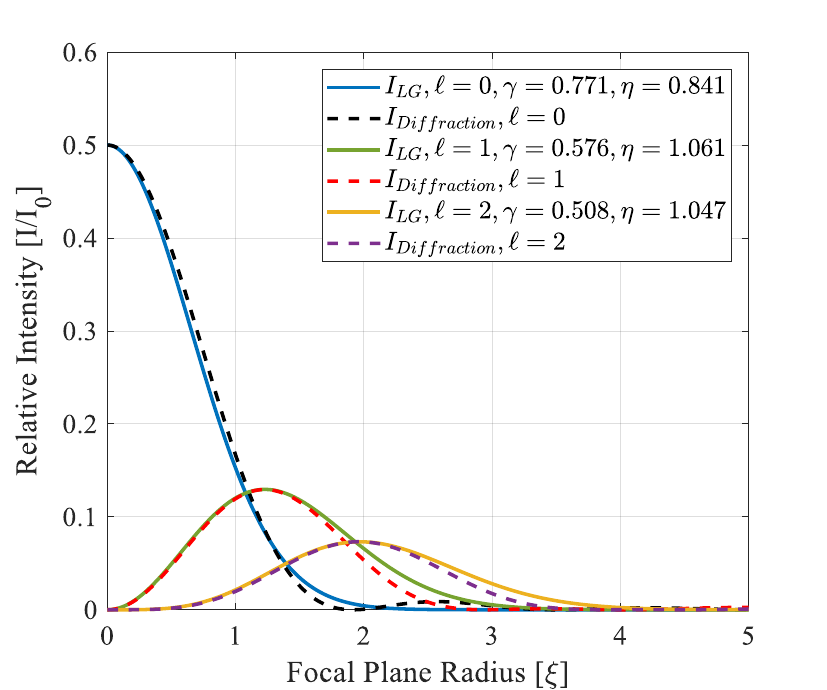}
\caption{Laguerre-Gaussian beams fitted to the exact diffraction results (dashed) for OAM beams driven by flat-top near-field beams for $\ell=0$, $\ell=1$, and $\ell=2$ topological charges.}
\end{figure}
We have additionally included in Table 1 and Fig.4 the fitting parameters of the $\ell=0$ LG mode. These parameters were found by matching the beam waist $w_0$ of the LG mode to the $1/e^2$ value of the diffraction result. This is a useful result for those wishing to accurately model the Gaussian focus of a HPL given the focal length and radius of the flat top beam in the near-field. 

\section{Super-Gaussian OAM beams}
In reality, HPL's have neither a flat-top or a Gaussian near field beam. Rather, a convolution of the two resulting in the well known super-Gaussian beam \cite{Grow:06}. The near-field can take the following form,
\begin{equation} \label{eq22}
    U(r)=U_0\sqrt{\frac{2^{1/n}n}{2\Gamma(1/n)}}exp\left[-\left(\frac{r}{R_0}\right)^{2n}\right]
\end{equation}where $\Gamma(x)$ is the gamma function. The normalization factor ensures that the total power in the near-field is the same as in the previous sections, and as such, the definition of the beam radius $R_0$ remains the same. This is clear to see as if one sets the super-Gaussian parameter $n$ equal to one, we retrieve the Gaussian near-field, and if we set $n=\infty$, then we retrieve the flat-top near-field. Eq.\ref{eq22} is therefore a generalization of the previous near-fields. In practice, the near-field of the HPL may not be circularly symmetric and could be elliptical or rectangular depending on the amplifier geometry. We however limit the study to circularly symmetric super-Gaussian beams such that we can continue to use the Hankel transform in Eq. \ref{eq5}. In general, there is no known analytic solutions to the Hankel transform of Eq.\ref{eq22} so we rely on a numerical study in this section. Based on our previous methods, we will assume the far-field radius can be also be normalized to the same beam waist value, $\xi=\rho/w_0$. 

A typical HPL may have a super-Gaussian near field on the order of $n=2$ to $5$, the far-fields of which are given in Figs.5 and 6 for the first two $\ell$ modes.  We also include the results for a near-field beam with $n=1$ (Gaussian driver), and for $n=\infty$ (flat-top driver) for comparison. Again we assume no power losses from the focussing optic, therefore the power in the focal spot is constant from one OAM beam to the next. It is clear to see that as $n$ increases, the focal spot intensity profile quickly converges to the result of an ideal flat-top beam ($n=\infty$). From Figs. 5 and 6, it is also clear to see that super-Gaussian lasers with $n>3$ can be well represented by a flat-top beam.  

\begin{figure}[ht!] \label{fig5}
\centering\includegraphics{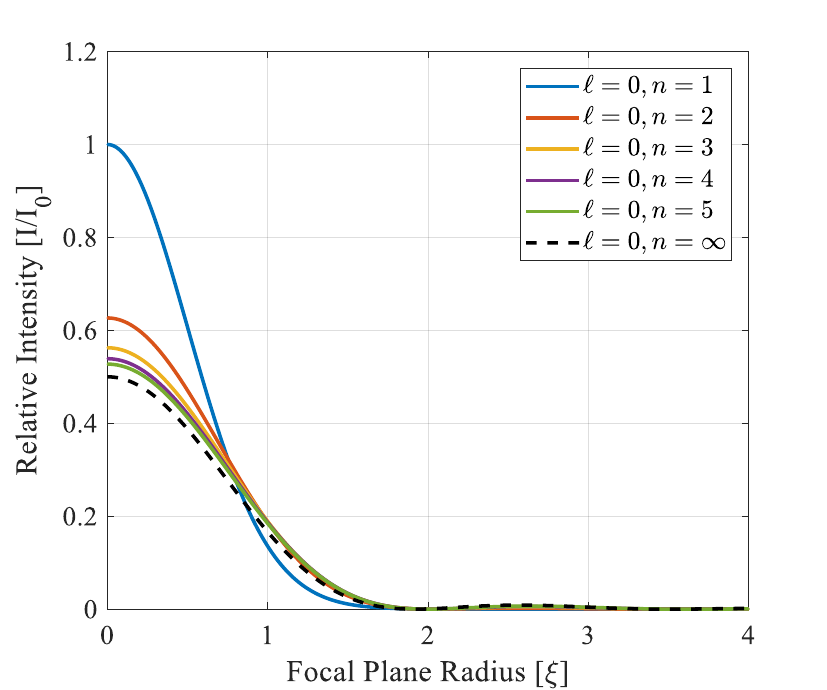}
\caption{The first 5 super-Gaussian focal spots for an $\ell=0$ beam. The dashed line represents a flat-top driver beam. The blue line corresponds to a Gaussian driver.}
\end{figure}
\begin{figure}[ht!] \label{fig6}
\centering\includegraphics{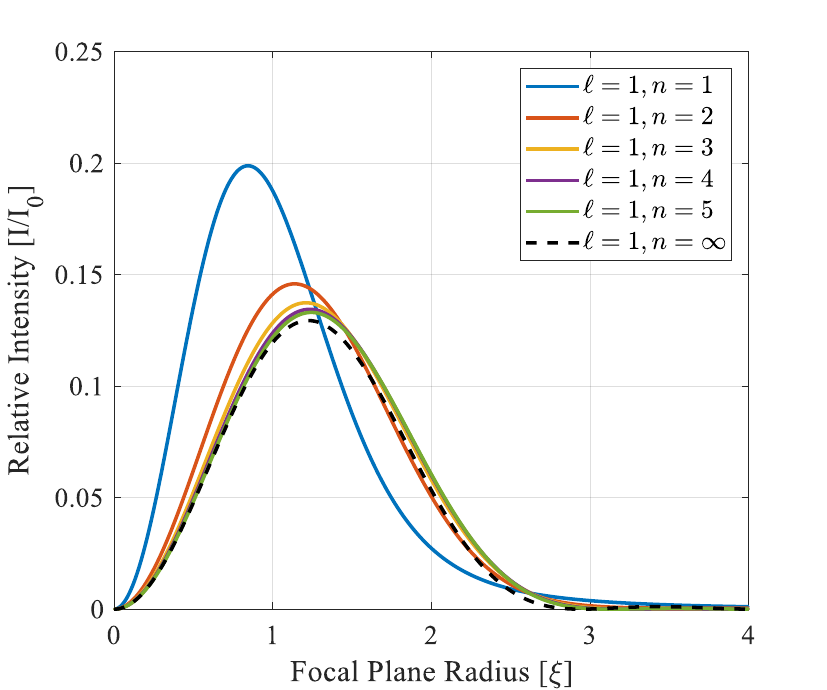}
\caption{The first 5 super-Gaussian focal spots for an $\ell=1$ beam. The dashed line represents a flat-top driver beam. The blue line corresponds to a Gaussian driver}
\end{figure}

As before, we want to find an optimal LG mode to simply represent each of the OAM beams at focus for various $\ell$ and $n$ values. Using the relations given previously in Eq's.\ref{eq13} and \ref{eq15}, we can find the suitable values of $\gamma$ and $\eta$ for the LG modes. We have removed the subscript from the $\eta$ and $\gamma$ variables as both the Gaussian and flat-top driven OAM beams can be contained within the super-Gaussian set. These values have been tabulated in Tables 2 and 3, respectively, for the first 6 topological charges and for more useful $n$ values. 

\begin{table}\label{table2}
\centering 
\caption{
		Optimal beam waist ratios $\gamma$ of the first 6 $\ell$ numbers for Gaussian ($n=1$), flat-top ($n=\infty$), and useful super-Gaussian mode numbers. 
	}
	
		\begin{tabular}{c c c c c c c c}
		\hline
			\textrm{$\ell \char`\\ n$}&
			\textrm{$1$}&
			\textrm{$2$}&
			\textrm{$3$}&
			\textrm{$4$}&
			\textrm{$5$}&
			\textrm{$10$}&
			\textrm{$\infty$}\\
			\hline
			0 & 1.000 & 0.797 & 0.766 & 0.758 & 0.756 & 0.759 & 0.771 \\
			1 & 0.837 & 0.622 & 0.584 & 0.574 & 0.569 & 0.568 & 0.576 \\
			2 & 0.746 & 0.554 & 0.519 & 0.508 & 0.504 & 0.503 & 0.508 \\
			3 & 0.680 & 0.508 & 0.477 & 0.466 & 0.460 & 0.459 & 0.465 \\
			4 & 0.629 & 0.475 & 0.443 & 0.434 & 0.430 & 0.426 & 0.432 \\
		    5 & 0.583 & 0.447 & 0.418 & 0.409 & 0.404 & 0.401 & 0.407 \\
			\hline
		\end{tabular}
\end{table}

\begin{table}\label{table3}
\centering 
\caption{
		Optimal conversion efficiencies $\eta$ of the first 6 $\ell$ numbers for Gaussian ($n=1$), flat-top ($n=\infty$), and useful super-Gaussian mode numbers. 
	}
	
		\begin{tabular}{c c c c c c c c}
		\hline
			\textrm{$\ell \char`\\ n$}&
			\textrm{$1$}&
			\textrm{$2$}&
			\textrm{$3$}&
			\textrm{$4$}&
			\textrm{$5$}&
			\textrm{$10$}&
			\textrm{$\infty$}\\
			\hline
			0 & 1.000 & 0.986 & 0.960 & 0.939 & 0.924 & 0.885 & 0.841 \\
			1 & 0.772 & 1.023 & 1.094 & 1.110 & 1.116 & 1.101 & 1.061 \\
			2 & 0.589 & 0.892 & 1.001 & 1.046 & 1.062 & 1.070 & 1.047 \\
			3 & 0.472 & 0.778 & 0.908 & 0.970 & 1.005 & 1.035 & 1.010 \\
			4 & 0.391 & 0.680 & 0.828 & 0.901 & 0.939 & 0.996 & 0.980 \\
		    5 & 0.338 & 0.603 & 0.756 & 0.837 & 0.882 & 0.952 & 0.946 \\
			\hline
		\end{tabular}
\end{table}

An example of optimized LG modes fitted to $n=2$ super-Gaussian modes are given in Fig.5. We find remarkable agreement between the fitted result and diffraction theory. From Table 3 we note that there are a few LG modes that have a scaling factor close to unity, for instance the mode that best fits the $\ell = 2$, $n=3$ diffraction result has a scaling factor of $\eta = 1.001$. This is in contrast to the modes that were fitted to the Gaussian near-field focal spots that had scaling factors less than 0.5 for $\ell>2$. This method of LG fitting is therefore practical for super-Gaussian and flat-top driven OAM beams whereas the method outlined in \cite{Longman:17} may be more useful for describing Gaussian driven OAM beams where accurate representation of power in the focal spot is of concern. Otherwise, we find the present method to be superior for describing the peak intensity and general shape of the OAM beam at focus.   
\begin{figure}[ht!] \label{figure7}
\centering\includegraphics{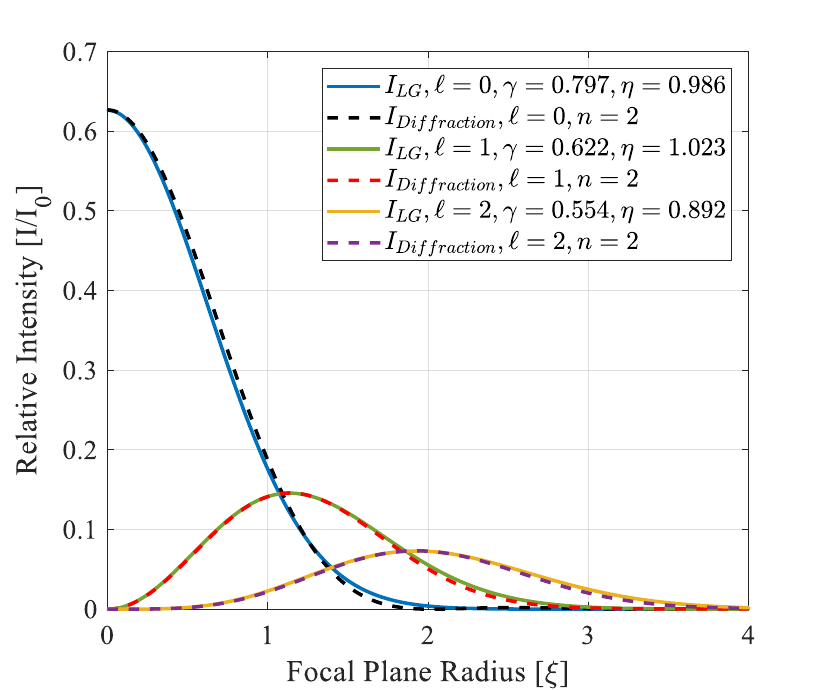}
\caption{Laguerre-Gaussian beams fitted to the exact diffraction result (dashed) for OAM beams driven by a $n=2$ super-Gaussian near-field beam for $\ell=0$, $\ell=1$, and $\ell=2$ topological charges.}
\end{figure}

\section{Conclusion}
In this article we have explored the scalar diffraction theory of beams carrying OAM from various high powered laser near-fields including Gaussian, super-Gaussian and flat-top beams. We have derived generalizations of the Gaussian and Airy focal spots that include OAM and found that there is an enhancement between the relative intensities of the $\ell>3$ and $\ell=0$ modes for a flat-top driven OAM mode as compared to a Gaussian driven one. We attribute this increase in relative intensity to the overlap of the OAM donut mode with the diffractive rings from the circular aperture function and due to the increased power at larger radii in the starting and diffracted beams leading to better matching to higher $\ell$ modes with larger radii. 

Laguerre Gaussian functions were fitted to each of the far-field intensity distributions to aid in the accurate numerical and analytic modelling of high-power OAM beams, showing that one can represent the intensity profile and the subsequent electric and magnetic fields of a high intensity OAM focal spot to a high degree of accuracy that well represents the diffraction theory. Additionally, the fitting parameters for the $\ell=0$ mode have been included for the best fit of a Gaussian profile to the Airy focal spot. 

This work should allow optimum modelling of both OAM beams and non-OAM beam fits to focal spots of high-power lasers  

\section*{Funding}
This work was funded by the Natural Sciences and Engineering Research Council of Canada (NSERC) research grant number RGPIN-2019-05013.

%%%%%%%%%%%%%%%%%%%%%%% References %%%%%%%%%%%%%%%%%%%%%%%%%

%%%%%%%%%% If using BibTeX:
\bibliography{references}

\end{document}